\documentclass[11pt]{article}
\usepackage{epsf,a4wide,amssymb,cite}
\makeatletter

\@addtoreset{equation}{section}
\def\section{\@startsection {section}{1}{\z@}{-3.5ex plus -1ex minus
 -.2ex}{2.3ex plus .2ex}{\large\bf\centering}}
\def\section{\@startsection {section}{1}{\z@}{-0.5ex plus -1ex minus
 -.2ex}{0.5ex plus .2ex}{\bf}}
\def\subsection{\@startsection{subsection}{2}{\z@}{-3.25ex plus%
 -1ex minus -.2ex}{1.5ex plus .2ex}{\bf}}
\def\subsection{\@startsection{subsection}{2}{\z@}{-0.25ex plus%
 -1ex minus -.2ex}{0.5ex plus .2ex}{\bf}}
\def\subsubsection{\@startsection{subsubsection}{3}{\z@}{-3.25ex plus%
 -1ex minus -.2ex}{1.5ex plus .2ex}{\sl}}
\makeatother

\newcommand{\be}{\begin{equation}}
\newcommand{\ee}{\end{equation}}
\newcommand{\bea}{\begin{eqnarray}}
\newcommand{\eea}{\end{eqnarray}}
\newcommand{\ba}{\begin{array}}
\newcommand{\ea}{\end{array}}
\newcommand{\nn}{\nonumber}

\def\ds{\displaystyle}

\def\blank#1{}

\def\cev#1.{{\langle{#1}|}}

\def\fract#1#2{{\textstyle\frac{#1}{#2}}}

\def\half{{\textstyle \frac 12}}
\def\haf#1{\half{(#1)}}
\def\hf#1{{\textstyle \frac {#1}2}}
\def\hf#1{{\ds\frac {#1}2}}

\def\is#1{{|\,#1\,\rangle\!\rangle}}
\def\no{{{:}}}

\newcommand{\One}{{\hbox{{\rm 1{\hbox to 1.5pt{\hss\rm1}}}}}}

\def\qt{{\textstyle \frac 14}}

\def\tq{{\tilde q}}
\def\TT{{\tilde\tau}}
\def\tn{{\tilde\nu}}

\def\vec#1.{{|{#1}\rangle}}

\newcommand{\vev}[1]{{\langle #1 \rangle}}

\def\vf{{\ensuremath{{\vec\rm free.}}}}
\def\vfp{{\ensuremath{{\vec\rm free'.}}}}
\def\vfs{{\ensuremath{{\vec\rm same.}}}}
\def\vfo{{\ensuremath{{\vec\rm {opp.}.}}}}

\newcommand{\vp}{{\varphi}}

\def\x{{\xi}}

\def\y{{\eta}}

\newcommand{\z}{\zeta}
\newcommand{\bz}{{\bar\zeta}}

\begin{document}
\parindent 12pt
\parindent 0pt
\parskip 9pt

{
\parskip 0pt
\newpage
\begin{titlepage}
\begin{flushright}
KCL-MTH-00-06\\
hep-th/0002218n\\
%Draft 4
Feb 25, 2000\\[3cm]
\end{flushright}
\begin{center}
{\Large{\bf
On the boundary Ising model with disorder operators
}}\\[1.4cm]
{\large G.M.T. Watts%
\footnote{e-mail: gmtw@mth.kcl.ac.uk}%
}\\[8mm]
{\em Department of Mathematics, King's College London,}\\
{\em Strand, London, WC2R 2LS, U.K.}\\[5mm]

{\bf{ABSTRACT}}
\end{center}
\begin{quote}
We extend the well-known method of calculating bulk correlation
functions of the conformal Ising model via bosonisation to situations
with boundaries.  
Oshikawa and Affleck have found the boundary states of two decoupled
Ising models in terms of the orbifold of a single free boson
compactified on a circle of radius $r{=}1$;
we adapt their results to include disorder operators.
Using these boundary states we calculate the expectation value of a
single disorder field on a cylinder with free boundary conditions and
show that in the appropriate limits we recover the standard and
frustrated partition functions. We also show how to calculate Ising
correlation functions on the upper half plane. 
\end{quote}
\vfill
\end{titlepage}
}

\section{Introduction}
\label{sec:one}
\setcounter{footnote}{0}

There is a well-known construction of correlation functions in two
decoupled critical Ising models as correlation
functions in a $c{=}1$ free boson theory \cite{FSZu1,ybk}.
A natural question to ask is how to extend this to the boundary model
and how to extend the analysis of Cardy and Lewellen \cite{C,CL} to
include non-local bulk fields such as the Ising disorder field.

We first recall the Ising model field content. 
The maximal set of primary fields one normally considers are
the identity field $\One$ of weight $(0,0)$;
the spin field $\sigma$ and disorder field $\mu$ of weights
$(1/16,1/16)$; the energy operator $\epsilon$ of weight $(1/2,1/2)$;
and the fermion fields $\psi$ and $\bar\psi$ of weights $(1/2,0)$ and
$(0,1/2)$ respectively.
These fields are not mutually local, and the three maximal sets of
mutually local fields are
\[
\{ \One,\sigma,\epsilon \} \;,\;
\{ \One,\mu,   \epsilon \} \;,\;
\{ \One,\psi, \bar\psi, \epsilon \}
\;.
\]
We shall take the point of view that our primary fields are the local
set $\One$, $\sigma$ and $\epsilon$ and we can then add the disorder
field $\mu$ at the expense of introducing disorder lines joining pairs
of disorder fields. 
The effect of the disorder lines is that the correlation function
changes sign when a spin field passes through a disorder line;
without these disorder lines the correlation
functions containing both spin and disorder fields would be
double-valued. 
We also recall that 
bulk correlation functions are related by duality, under which
$ \sigma \leftrightarrow \mu $, $\epsilon \leftrightarrow -\epsilon$.

A conformally-invariant boundary condition on a conformal field theory
the upper half plane defines a set of correlation functions of the
bulk fields satisfying $ T(x) = \bar T(x)$ on the $x$--axis.
The boundary theory of the local field theory containing $\sigma$ and
$\epsilon$ was investigated by Cardy \cite{C} who found that
(under certain assumptions such as the uniqueness of the vacuum) there
are three such boundary conditions, denoted by `$+$', `$-$' and `$f$',
which have the interpretation that the spin variable in the lattice
realisation is fixed up, fixed down and free, respectively.

The fields on the boundary can be classified into primary and
descendent fields.
Under Cardy's assumptions, one finds that the identity is the only
primary field on the $\pm$ boundaries, whereas on the free b.c.\ there
is a non-trivial primary field $\sigma^B(x)$ of weight $1/2$ which has
the interpretation of the boundary spin field.  
One can further consider fields $\psi^{(\alpha\beta)}$ which interpolate
different b.c.'s\ $\alpha$ and $\beta$, and which again fall into
primary and descendent fields;
$\psi^{(\pm \mp)}$ have weight 1/2 and 
$\psi^{(\pm f)}, \psi^{(f \pm)}$ have weight 1/16.

To complete the picture one needs the structure constants in the
bulk-boundary opes (expressing the expansion of a bulk field in
boundary fields) and the opes of boundary fields.
%The general expressions for these structure constants in A and D type
%minimal models have been given
%in \cite{Runkel}; 
For the Ising model these were found by Cardy and Lewellen \cite{CL},
and the bulk-boundary opes take the form 
\be
\begin{array}{rclrcl}

  \left.  \sigma(z,\bar z) \right|_{\pm}
& \sim 
&  {}^{(\pm)} B_\sigma^1 \cdot     |2y|^{-1/8}
 
\qquad \qquad

& \left. \epsilon(z,\bar z) \right|_{\alpha}
& \sim 
& {}^{(\alpha)} B_\epsilon^1 \cdot |2y|^{-1}

\\[2mm] 

  \left.  \sigma(z,\bar z) \right|_{f}
& \sim 
&  {}^{(f)} B_\sigma^{\sigma^B} \cdot  |2y|^{3/8}\, \sigma^B(x)

\end{array}
\label{eq:sigmabbope}
\ee
where
$ 
{}^{(\pm)} B_\sigma^1 = \pm 2^{1/4} ,\;
{}^{(f)} B_\sigma^{\sigma^B} = 2^{-1/4} ,\;
{}^{(\pm)} B_\epsilon^1 = 1 ,\;
{}^{(f)} B_\epsilon^1 = -1.
$

A natural question is how to include the bulk disorder field in this
analysis.
Roughly speaking, fixed and free b.c.'s are interchanged under
duality.
However, as noted in \cite{Miguel}, this leads to an
apparent contradiction between duality and the (conjectured)
$g$--theorem for boundary renormalisation group flows.
This conjecture states that the boundary entropy $g$ in a unitary theory
decreases along a renormalisation group flow \cite{ALud1}. 
The values for the free
and fixed b.c.'s are $g_{\rm free}=1$, $g_{\rm fixed}=2^{-1/2}$.
When the free b.c.'s are perturbed by a boundary magnetic field, there
is a flow to the fixed b.c. However, under duality there should be a
similar flow from fixed to free induced by the boundary disorder
field, which would apparently be forbidden by the $g$--theorem.
A resolution of this contradiction was obtained in \cite{Miguel2} in
which it was shown that a careful consideration of the two local field
theories leads one to the result that for the 
local theory containing $\{\One,\mu,\epsilon\}$ one has three
different boundary conditions, `fixed', `free+' and `free$-$' which
are dual to `free', `fixed+' and `fixed$-$'. 
In terms of these new b.c.'s, the flow is from `fixed' with $g=1$ to 
`fixed$\pm$' with $g=2^{-1/2}$, in agreement with the $g$--theorem.
Here we suggest another resolution adapted to our situation in which
we wish to remain within a theory containg the spin field.
As we see below, the boundary disorder field is locally identical to
the operator interpolating fixed $+$ and fixed $-$ b.c.'s, and so 
one can formulate a perturbation by the boundary disorder field only
in a Hilbert space describing the direct sum of $+$ and $-$ b.c.'s. 
Since the actual flow is now from a direct sum of $+$ and $-$ b.c.'s
with  $g \,{=}\, 2\cdot 2^{-1/2} \,{=}\, 2^{1/2}$ to free with
$g\,{=}1\,$, $g$ decreases along this flow, and again there is no
contradiction with the $g$--theorem. 

Using duality, the bulk-boundary ope of the disorder field in the
free b.c.\ should be dual to that of the spin field in fixed b.c.'s; 
i.e.\ in the free b.c.\ the bulk disorder field only couples to the
identity, 
\be
  \left. \mu(z,\bar z) \right|_{\rm free}
\sim
  {}^{(\rm free)}B_{\mu}^1 \, |2y|^{-1/8}
\;,\;\;
  {}^{(\rm free)}B_{\mu}^1 = \pm 2^{1/4}
\;,
\label{eq:mubbope1}
\ee
where the sign ambiguity reflects the overall ambiguity in correlation
functions involving the disorder field.

However, when a disorder field approaches a free boundary, it carries
its associated disorder line with it.
Since the leading term in the ope of the disorder field with the
boundary is the identity operator, this means that
a disorder line can end on a free boundary at an
operator of weight 0.
This explains the observation in \cite{Ruelle} that
the boundary state describing a circular boundary in the plane with
free b.c.\ with a disorder line is rotationally invariant -- one might
at first sight think that  the presence of the disorder line ending on
the boundary would lead to a dependence on the position of the
disorder line on the boundary, but since it couples to  field of
weight 0 this is not the case. Alternatively, the disorder line simply
represents a cut in the value of correlation functions and has no
physical content, it does not break the rotational symmetry of the
correpsonding boundary state.

We shall see this directly in section
\ref{sec:cyl} where we construct the one-point function of the bulk
disorder field on the cylinder with free b.c.'s which interpolates the
frustrated and unfrustrated partition functions as the disorder field
crosses the cylinder.

In fixed b.c.'s conversely, the bulk disorder field will only couple
to a boundary field of weight 1/2, the boundary disorder field (also
called the boundary freedom field \cite{Miguel})
$\mu^B(x)$, 
\be
  \left. \mu(z,\bar z) \right|_{\rm fixed}
\sim
  {}^{(\rm fixed)}B_{\mu}^{\mu^B} \, |2y|^{3/8} \cdot \mu^B(x)
\;,\;\;
  {}^{(\rm fixed)}B_{\mu}^{\mu^B} = 2^{-1/4}
\;,
\label{eq:mubbope2}
\ee
We have denoted the b.c.\ by `fixed' rather than by `$\pm$' 
since the presence of the disorder fields means that it is not
possible to unambiguously identify a fixed b.c.\ as either `$+$' or 
`$-$'. 
The ability to move a disorder line across a
boundary at no cost (other than a sign change when it passes
through a spin field) is also true for the fixed b.c., thereby
changing the b.c. from $+$ to $-$ or vice versa. 
Using this property we see that the boundary disorder field on a fixed
b.c.\ can be identified with a $(\pm \mp)$ boundary condition changing
operator connected to a disorder line, as shown in figure 1

{
%\begin{figure}[ht]
\[
\ba{ccc}
\epsfxsize=.9\linewidth
\epsfbox{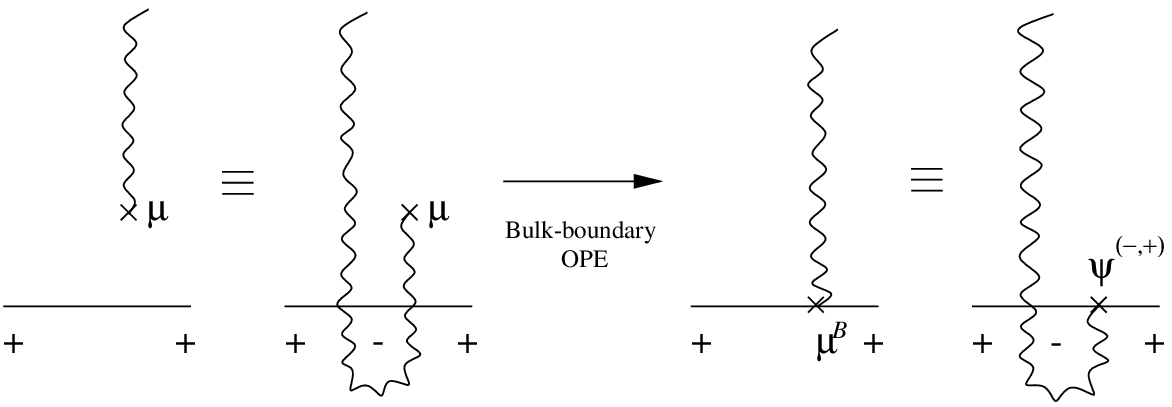}
\\[2mm]
\parbox[c]{.6\linewidth}
{\raggedright
Figure 1: the equivalence of the boundary disorder field and the
boundary changing operator}
\ea
\]
}

One can `pull' the disorder line attached to the bulk disorder field
down and to the left until it crosses the edge of the upper-half-plane
leaving the bulk disorder field attached to a disorder line that ends
directly on the boundary at a field of weight 0.
When the bulk field then approaches the boundary, this disorder line
shrinks and locally the leading term in the bulk-boundary ope is the
boundary-condition-changing operator interpolating $\pm$ b.c.'s.
This shows that the boundary disorder field is locally identical to a
boundary-condition changing operator, the only difference being the
position of disorder lines.

%\newpage

\section{Bosonisation formulae}
\label{sec:bf}

As explained in \cite{G}, one can realise two independent copies of
the Ising model with local fields $\{1,\sigma,\epsilon\}$ as the
$\mathbb Z_2$
orbifold of a single free boson compactified on a circle of radius
$r{=}1$. If we label the fields in the two Ising models by 1 and 2, this
is generated by
\be
\{\, \One, 
  \,\sigma_1, 
  \,\epsilon_1, 
  \,\sigma_2, 
  \,\epsilon_2\,\}
\;,
\label{eq:A}
\ee
and is a local field theory;
we shall call this model A.
One can consider a different $c=1$ theory containing (amongst others)
the following alternative mutually local set of fields,
\be
  \{
  \, \One,
  \, \sigma_1\sigma_2, 
  \,\mu_1\mu_2, 
  \,\epsilon_1{+}\epsilon_2, 
  \,\epsilon_1\epsilon_2
\,\}
\;,
\label{eq:B}
\ee
which we shall call model B.
This is the $\mathbb Z_2$ orbifold of model A, i.e.\ the un-orbifolded
free boson and contains only symmetric combinations of the fields in
the two Ising models, although not all -- for example it does not
contain $\sigma_1+\sigma_2$ which would be non-local with respect to
$\mu_1\mu_2$. 

The advantage of considering model B is that it enables one to
construct correlation functions $\vev{\,{\cal O}\,}$
in a single Ising model as the square root of the correlation function
$\vev{\,{\cal O}_1{\cal O}_2\,}$ in a free boson theory, as explained
in \cite{FSZu1}.
The correlation functions in model B are single-valued, but taking the
square--root leads to the (possibly) multi-valued correlation functions
in the Ising model.

The bosonisation formulae on the plane are \cite{FSZu1}
\be
\begin{array}{rclrcl}

    \sigma_1  \sigma_2
&{=}&   
    \sqrt 2\; \no\cos \half \vp \no 
\;,

\qquad

&   \mu_1  \mu_2
&{=}&   \!\!\!
    \sqrt 2\; \no\sin  \half \vp \no 
\;,

\\[2mm]

    \epsilon_1  \epsilon_2
&{=}&
  -\;\no \partial\vp\,\bar\partial\vp \no 
\;,

&   \epsilon_1(z,\bar z)  \,{+} \,  \epsilon_2(z,\bar z)
&{=}&  
    2\; \no \cos \vp \no 
\;,

\end{array}
\label{eq:bose}
\ee
Note that the field representing $\mu_1\mu_2$ is absent from the
orbifold model of \cite{OA}.

We now consider this model on a cylinder of 
length $L$, circumference $R$, which we take to be the rectangle in
the complex $z$ plane with vertices $0, R, iL, R+iL$.
This can be mapped to an annulus in the $\z$ plane by
$\z = \exp( - 2 \pi i z/ R)$.
In this form, the boundary conditions on the two ends of the cylinder
are represented by boundary states in the full bulk theory.

All the possible boundary states for the model A were found by
Oshikawa and Affleck in \cite{OA}.
To use their results, we first give our conventions.
We take the mode expansion of $\vp(\z,\bz)$ on the plane to be
\[
  \vp(\z,\bz)
= q  - i p \ln(\z \bz) - i w  \ln(\z/\bz)
+ i \sum_{n\neq 0} ( \frac{a_n}n \z^{-n} + \frac{\bar a_n}n\bz^{-n} )
\;,
\]
where $w$ is the winding number, $[q,p]=i$ and 
$[a_m,a_n] = [\bar a_m,\bar a_n] = m \delta_{m,-n}$.
With this choice of $\vp$, compactification on a circle of radius 1 in
the sense of \cite{G} means we identify $\vp \cong \vp + 4\pi$, and hence
$w$ takes integer and $p$ half-integer values (n.b.~this is different
to \cite{ybk}).
For the boundary states we consider, we only need the Dirichlet-type
Ishibashi states of zero winding number, viz.
\be
  \is{k}
= \exp( \sum_{n>0} \frac{a_{-n}\bar a_{-n}}n  )\vec k.
\;,
\label{eq:is1}
\ee
which satisfy
\be
  (a_n - \bar a_{-n}) \is k = 0
\;.
\label{eq:aba}
\ee 
The boundary states in the orbifold model also contain states in the
twisted sector, but these will play no role for us.

Although the space of states in models A and B differ, one of the
boundary conditions of \cite{OA} is common to the two models, and that
is the condition that the spins on both models are `free'.
We denote this by \vf\ and it is given by
\be
 \vf
= \sum_{k\in\mathbb Z} (-1)^k \, \is k
\;.
\label{eq:free}
\ee
To check that this does indeed correspond to free boundary conditions,
one can calculate the partition function for the system with this
boundary state on the two ends.
Given the cylinder Hamiltonian
\[
  H(R) 
= \frac{2\pi}R( L_0 + \bar L_0 - 1/12)
\,
\]
the partition function with this `free'
boundary condition on the both ends of the cylinder is
\bea
    Z_{\rm free,free}
&=& \cev \rm free. \, e^{-L H(R)}\, \vec \rm free.
%\nn\\
%&=& 
 = \sum_{k\in\mathbb Z} \tq^{k^2/2} / \eta(\TT)
%\nn\\
%&=& 
 = \theta_3(0|\TT)/\eta(\TT)
%\nn\\
%&=&
 = \theta_3(0|\tau)/\eta(\tau)
\nn\\
&=& ( \chi_0(q) + \chi_{1/2}(q) )^2
%\nn\\
%&=&
 = ( Z^{\rm Ising}_{\rm free,free} )^2
\;,
\label{eq:Zff}
\eea
where $\tq=\exp(-4\pi L/R)$, $q= \exp(-\pi R/L)$,
$\TT = 2i L/R$, $\tau=iR/(2L)$, 
$\chi_h(q)$ is the 
character of the $c=1/2$ Virasoro algebra representation of weight
$h$,
and the definitions and properties of the $\theta$ functions can be
found in e.g.\ chap.\ 10 of \cite{ybk}.

Hence the state \vf\ naturally describes free boundary
conditions with no disorder lines (or an even number of disorder
lines) ending on the boundary.
To find the boundary state \vfp\ for the system with an odd
number of 
disorder lines, we can use the fact that a bulk disorder operator
carries with it a disorder line and has the bulk--boundary ope
\[
 \left. \mu(x+iy) \right|_{\rm free}
\sim 2^{1/4}\, (2y)^{-1/8}\, \One + \ldots
\;.
\]
Hence we can calculate \vfp\ by acting with the bosonic expression for
$\mu_1\mu_2$ on \vf.

Using the standard expression for the normal ordered vertex operator
on the cylinder expressed in terms of the field on the plane,
\[
%  \exp( i \alpha \vp(z,\bar z))
\no  e^{ i \alpha \vp_{\rm cyl.}(z,\bar z) } \no
=    (\fract{2\pi}R)^{\alpha^2}
  \, |\z|^{\alpha^2}
%  \, \exp(i \alpha\phi_<(\z,\bz))
  \, e^{i \alpha\vp_<(\z,\bz)}
  \, e^{i\alpha q}
  \, |\z|^{2\alpha p}
%  \, \exp(i \alpha\phi_>(\z,\bz))
  \, e^{i \alpha\vp_>(\z,\bz)}
\;,
\]
where $\z = \exp(-2\pi i z/R)$ and
\[
\textstyle
  \vp_{{>}\atop{<}}(\z,\bz)
= i\sum_{n{> \atop <}0} (\frac{a_n}n \z^{-n} + \frac{\bar a_n}n \bz^{-n})
\;,
\]
one easily finds that
\be
%  \exp( i\alpha\vp(z,\bar z)) 
\no  e^{ i\alpha\vp_{\rm cyl.}(z,\bar z) } \no
  \is k
%=    (\fract{2\pi}R)^{\alpha^2}
%  \, ( |\z| - |\z|^{-1} )^{-\alpha^2}
=    (\fract R\pi\, {\rm sinh} \fract{2\pi y}R )^{-\alpha^2}
  \, |\z|^{2\alpha p}
%  \, \exp(i \alpha( \phi_<(\z,\bz) - \vp_<(1/\z,1/\bz)))
  \, e^{ i \alpha( \vp_<(\z,\bz) - \vp_<(1/\z,1/\bz)) }
  \, \is{k+\alpha}
\;.
\label{eq:vois}
\ee
Therefore, using the bosonisation formulae (\ref{eq:bose}), we find that
\be
  \mu_1\mu_2(x+iy) \vf
= - i \sqrt 2 ( 2 y )^{-1/4} 
  \sum_{k\in\mathbb Z} (-1)^k \, \is{k+1/2} 
+ O(y^{3/4})
\;,
\ee
and so the boundary state \vfp\ of a free boundary condition with an 
odd number of disorder lines is given by
\be
  \vfp
= -i \sum_{k\in\mathbb Z}(-1)^k\,\is{k+1/2}
% = \sum_{r\in\mathbb Z+1/2} (-i)^r \, \is r
\;.
\label{eq:vfp}
\ee

As a check we can calculate the partition function for the Ising
model with a disorder line ending on each end, 
\bea
    Z_{\rm free',free'}
&\!\!=\!\!& \cev \rm free'. \, e^{-L H(R)}\, \vfp
%\nn\\
%&=& 
 \;= \! \sum_{r\in\mathbb Z+1/2} \tq^{r^2/2} / \eta(\TT)
%\nn\\
%&=& 
 \,=\, \theta_2(0|\TT)/\eta(\TT)
%\nn\\
%&=&
 \,=\, \theta_4(0|\tau)/\eta(\tau)
\nn\\
&\!\!=\!\!& ( \chi_0(q) - \chi_{1/2}(q) )^2
%\nn\\
%&=&
 \,=\, ( Z^{\rm Ising}_{\rm free',free'} )^2
\;,
\label{eq:Zff2}
\eea
i.e.\ the square of the frustrated partition function \cite{Ruelle} as
should be the case.

As a final check, 
\be
  Z_{\rm free,free'}
= \cev{\rm free}. e^{-L H(R)} \vfp
= 0
\;,
\ee
reflecting the fact that there are no configurations with a single
disorder line ending on one end of the cylinder and no disorder
fields.

Having found the boundary states for the two Ising models in (free)
and (free'), we must complete the discussion by considering the case
of fixed boundary conditions. 
In model A, all possible combinations of up and down conditions on the
two models are possible, and all such boundary states have been found
in \cite{OA}. 
In model B however, the spin fields only appear in the symmetric
combination $(\sigma_1\sigma_2)$, and so one can only possibly
distinguish the relative signs of the two spins. This means that the
possible boundary conditions in model B are `fixed same' and
`fixed opposite', giving the relative signs of the two spins.
If we denote the corresponding boundary states by \vfs\ and \vfo, they
are given in terms of Oshikawa's and Affleck's boundary states as
\be
\begin{array}{rclcl}
    \vfs 
&=& \ds \frac{ \vec ++. + \vec --.}{\sqrt 2}
&=& \sum_{k\in \mathbb Z} \is{ k/2}
\\
    \vfo
&=& \ds \frac{ \vec +-. + \vec -+.}{\sqrt 2}
&=& \sum_{k\in \mathbb Z} (-1)^k \is{ k/2}

\end{array}
\label{eq:vfsvfo}
\ee
Using (\ref{eq:bose}) and (\ref{eq:vois}) one finds that
\[
\begin{array}{rcl}
       \sigma_1\sigma_2(x+iy)\,\, \vfs
&=& + \sqrt2 \, (2y)^{-1/4} \, \vfs \;+\; O(y^{3/4}) 
\;,
\\[2mm]
       \sigma_1\sigma_2(x+iy)\,\, \vfo
&=& - \sqrt2 \, (2y)^{-1/4} \, \vfo \;+\; O(y^{3/4}) 
\;.
\end{array}
\]

As a final comment, we note how the fixed and free boundary states are
related by duality.
If $V\,\vp\,V = -\vp$, then duality $\vp\mapsto\pi-\vp$ is implemented
by the operator
$ D = V\, e^{i\pi p}$. Hence we have
\[
  D \is k = e^{i\pi k}\is{-k}
\;,\;\;\;\;
  D\,\vf  =   \frac{\vfs + \vfo}{\sqrt 2}
\;,\;\;\;\;
  D\,\vfp =   \frac{\vfs - \vfo}{\sqrt 2}
\;.
\]

% \newpage
\section{Cylinder correlation functions}
\label{sec:cyl}

We can also use these bosonisation formulae and boundary states to
calculate correlation functions of fields on the cylinder.
One interesting case to consider is the expectation value of a single
disorder field with free b.c.'s on the two sides of the cylinder. 

To recall, the cylinder is given by the rectangle in the upper half
$z$ plane with vertices $0, R, iL, R + iL$, with the vertical edges 
${\rm Re}(z){=}0,R$ identified.
We must also choose on which end of the cylinder the disorder line
attached to the bulk field ends. We choose it to end on the top edge
${\rm Im}(z){=}L$.

Using standard free field techniques, we can explicitly evaluate
\bea
    \vev{\, \mu_1\mu_2(iy)\,}
&=& \cev {\rm free}'. \, e^{-L H(R)} \, 
    (\sqrt 2 \no \sin \half \vp(iy) \no) \,
    \vf
\nn\\
&=& \sqrt 2
    \, \frac{\theta_3(\tn/2|\TT)}{\eta(\TT)}
    \, \left[ \frac{1}{iR} 
              \frac{\theta'_1(0|\TT)}{\theta_1(\tn|\TT)}
       \right]^{1/4}
\nn\\
&=& \sqrt 2
    \, \frac{\theta_3(\nu/2|\tau)}{\eta(\tau)}
    \, \left[ \frac{1}{2L} 
              \frac{\theta'_1(0|\tau)}{\theta_1(\nu|\tau)}
       \right]^{1/4}
\;,
\eea
\nn\\
where $\nu=y/L$ and $\tn = -2iy/R$.
This gives the cylinder expectation value in the Ising model as

\be
  \vev{ \mu(iy) }_{{\rm free}',\rm free}
= 2^{1/4}\,
  \sqrt{ \frac{\theta_3( \frac\nu2|\tau) }{\eta(\tau)}  } \,
  \left| \frac{ \theta'_1(0|\tau)}{2 L\,\theta_1( \nu |\tau)} \right|^{1/8}
\;.
\label{eq:muvev}
\ee
This can also be found as the appropriate combination of the
two chiral blocks given in (12.109) %, p.~462 
of \cite{ybk}.

%Since the disorder field must be attached to a disorder line which can
%only end on a boundary, we should investigate on which boundary this is.

As the disorder field approaches the $x$--axis, $\nu\to 0$,
we find
\bea
  \vev{ \mu(iy) }_{\rm free,free}
&=&
  2^{1/4}\,
  (2\pi\nu)^{-1/8}\,
  \left(
  \sqrt{ \frac{\theta_3( 0 |\tau) }{\eta(\tau)} }  +  O(\nu)
  \right)
%\nn\\
%&=&  
\,=\,
   2^{1/4}\,
  (2y)^{-1/8}\,
  \left( \chi_{0} + \chi_{1/2} \right) + \ldots
\nn\\
&=&
   2^{1/4}\,
  (2y)^{-1/8}\,
  Z^{\rm Ising}_{\rm free,free} + \ldots
\;.
\eea
This is the exactly what we would have expected from the bulk-boundary
ope of the disorder field (\ref{eq:mubbope1}) at a free boundary.
Hence we see explicitly that as the disorder field approaches the
$x$--axis the disorder line shrinks to zero. 

However, if the disorder field approaches the other side, the disorder
line gets stretched across the cylinder. 
As it approaches the boundary ${\rm Im}(z){=}L$ the leading term in
the bulk-boundary ope will again be the identity operator, so that 
the leading behaviour of the correlation function should be the
frustrated partition function, i.e. the partition function including a
disorder line,
\[
  Z_{\rm free',free'}^{\rm Ising} 
= \chi_{0} - \chi_{1/2} 
;.
\]
Using (\ref{eq:muvev}) 
it is easy to check that this is in fact the case.  
Putting $ y = L {-} \tilde y$, $\tilde y = L v$,
$\nu = 1 - v$, we find
\bea
  \vev{ \mu(i(L - \tilde y)) }_{\rm free,free}
&=&
 2^{1/4}\,
  \sqrt{ \frac{\theta_4( \frac{v}2|\tau) }{\eta(\tau)}  } \,
  \left| 
  \frac{ \theta'_1(0|\tau)}{-2 L\,\theta_1( -v |\tau)} 
  \right|^{1/8}
\nn\\
&=&
  2^{1/4}\,
  (2Lv)^{-1/8}\,
  \left( 
  \sqrt{ \frac{\theta_4( 0 |\tau) }{\eta(\tau)} }  + O(v)
  \right)
\nn\\
&=&  
   2^{1/4}\,
  (2\tilde y)^{-1/8}\,
  \left( \chi_{0} - \chi_{1/2} \right) + \ldots
\;,
\eea 
as expected.
The two limits and the way the frustration line is stretched across
the cylinder are shown below in Figure 2.
{
\[
\ba{c}
\epsfxsize=.7\linewidth
\epsfbox{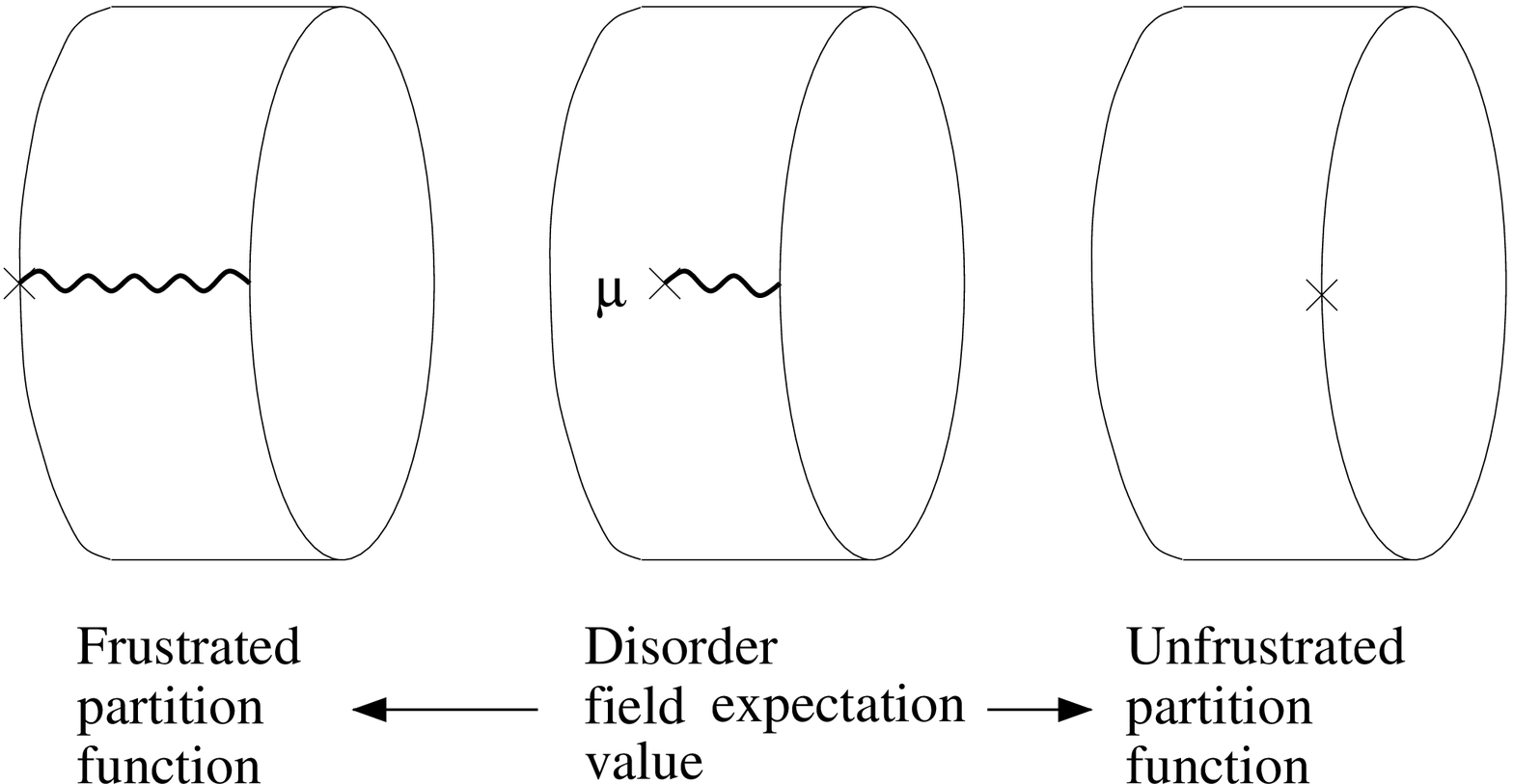}
\\
\hbox{Figure 2.}
\ea
\]
}
% \newpage
\newpage
\section{Correlation functions on the upper half plane}

We finish with some comments on the structure of Ising correlation
functions on the upper half plane (UHP).
From \cite{OA}, we can express the various boundary states in the B
model in terms of various combinations of Dirichlet boundary states
$\vec D(\vp_0).$, i.e.\ states for which $\vp$ takes the value $\vp_0$
on the boundary:
\[
\begin{array}{rclrcl}
  \vf  &=& \ds\frac{ \vec D(\pi). + \vec D(-\pi).}{\sqrt 2}
\qquad\qquad &
  \vfp &=& \ds\frac{ \vec D(\pi). - \vec D(-\pi).}{\sqrt 2}
\\
  \vfs &=& \vec D(0).
&
  \vfo &=& \vec D(2\pi).
\end{array}
\]  
Note that our field $\vp$ is twice that of \cite{OA} and hence the
values of $\vp_0$ are also twice theirs.

It is easy to express a free boson $\vp(\z,\bz)$ satisfying $\vp=\vp_0$
on the $x$--axis in terms of a chiral boson
$\phi(\z)$ as
\be
  \vp(\z,\bz) = \vp_0 + \phi(\z) - \phi(\bz)
\;.
\label{eq:vppp}
\ee
The only subtlety is that one cannot substitute this directly into
vertex operators as the result will be ill-defined.
We choose to define our normal ordering by
\be
  \left. \no\, \exp(i \alpha\,\vp(\z,\bz))\,\no\,\right|_{D(\vp_0)}
= (2 \y)^{-\alpha^2} 
  \, \no\, \exp( i\alpha( \vp_0 + \phi(\z) - \phi(\bar \z) ) \, \no
\;,
\ee
where $\z = \x + i\y$ with $\x, \y\in\mathbb R$.

If we also use the results of evaluating the
bulk--boundary opes, 
\[
  \begin{array}{rcl}
  \sqrt 2 (2\y)^{-1/4} \no \sin \haf{\phi(\z){-}\phi(\bz)} \no
& =
& \ds
  \frac{1}{\sqrt 2}
\, (2\y)^{3/4} 
\, \left[ \, i\partial\phi(\x) \;+\; O(\y) \,
  \right]
\;,
\\
  \sqrt 2 (2\y)^{-1/4} \no \cos \haf{\phi(\z){-}\phi(\bz)} \no
& = 
& \sqrt 2 
\, (2\y)^{-1/4}
\, \left[ 
  \, 1 \,+\, \half \y^2 \no(i\partial\phi(\x))^2\no \,+\, O(\y^3)
  \right]
\;,
\end{array}
\]
we can further find the expressions for the boundary spin and disorder fields.
We summarise the results in the following table of bosonisation formulae:
{\small\renewcommand{\arraystretch}{1.7}
\[
\begin{array}{cccc}
\hline

& {D(0)}
& {D(\pi)}
& {D(-\pi)}
\\\hline
\noalign{\vskip 2mm}

  \sigma_1\sigma_2
& ~\sqrt 2 (2\y)^{-\qt} \no \cos \hf{\phi(\z){-}\phi(\bz)} \no
& -\sqrt 2 (2\y)^{-\qt} \no \sin \hf{\phi(\z){-}\phi(\bz)} \no
& ~\sqrt 2 (2\y)^{-\qt} \no \sin \hf{\phi(\z){-}\phi(\bz)} \no
\\

  \mu_1\mu_2
& ~\sqrt 2 (2\y)^{-\qt} \no \sin \hf{\phi(\z){-}\phi(\bz)} \no
& ~\sqrt 2 (2\y)^{-\qt} \no \cos \hf{\phi(\z){-}\phi(\bz)} \no
& -\sqrt 2 (2\y)^{-\qt} \no \cos \hf{\phi(\z){-}\phi(\bz)} \no
\\

  \epsilon_1\epsilon_2
& \no \partial\phi(\z) (\partial\phi)(\bz) \no
& \no \partial\phi(\z) (\partial\phi)(\bz) \no
& \no \partial\phi(\z) (\partial\phi)(\bz) \no
\\

  \sigma^B_1\sigma^B_2
& \hbox{---}
& -i\partial\phi(\x)
& ~i\partial\phi(\x)
\\

  \mu^B_1\mu^B_2
& i\partial\phi(\x)
& \hbox{---}
& \hbox{---}

\\
\hline
\end{array}
\label{eq:ff2}
\]}%
Considering these formulae, we see that the only effect of taking the
particular linear combinations in the free and free$'$ b.c.'s
is to ensure that correlators including odd and even numbers of
disorder operators vanish respectively. 
As a result, we can safely evaluate correlators in the `generic' free
boundary conditions (i.e.\ not paying attention to the number of
disorder lines ending on the boundary) by using simply the
bosonisation formulae for $\vp_0=\pi$.
To summarise, we can calculate correlation functions in the UHP by
using expressions for $D(0)$ to obtain fixed b.c.'s in the Ising
models, and the expressions for $D(\pi)$ to obtain free b.c.'s.

Note that as we have defined them, the bosonisation formulae for $D(0)$
and $D(\pi)$ are not
exactly related by the duality $\vp \mapsto \pi - \vp$ which should
simply interchange the expressions for the spin and disorder fields
but instead differ by an irrelevant sign in the spin operator.

%%%%%% -- the plane --

As an example we use these formulae to calculate
a correlation function which would be hard to find
using standard conformal field theory techniques, being equivalent to
a five point chiral block.
Consider
the expectation value of a bulk spin field, a bulk disorder field and
a boundary disorder field in fixed boundary conditions.
Using the formulae for $D(0)$ we have
\bea
f(\z)\!&\!=\!&
\left(
   \vev{
\;\; \sigma(\z)
\;\; \mu(i)
\;\; \mu^B(1)
\;\;   }_{\rm fixed} 
\right)^2
 = 
%\nn\\[2pt]
%&\!=\!&
   \vev{
\;\; \sigma_1\sigma_2(\z)
\;\; \mu_1\mu_2(i)
\;\; \mu^B_1\mu^B_2(1)
\;\;   }
\nn\\[2pt]
&\!=\!& 
   2\,(2 \y)^{-1/4}\, (2)^{-1/4}\,
   \vev{ \,
   \no\cos(\half (\phi(\z) - \phi(\bz)))\no \;
   \;\no\sin(\half (\phi(i) - \phi(-i)))    \no \;
   \;i \partial\phi(0) 
    \, }
\nn\\
&\!=\!&
   -i \, 2^{-1/2}\,\y^{-1/4}
\nn\\
&&
    \left(
( \frac 1\z - \frac 1{\bar \z} + \frac 1i - \frac 1{-i} )
\!  \left[ \frac{ (i-\z)(i+\bar \z)}{(i-\bar \z)(i + \z)} \right]^{1/4}
-\;
( \frac 1\z - \frac 1{\bar \z} - \frac 1i + \frac 1{-i} )
\! \left[ \frac{ (i-\z)(i+\bar \z)}{(i-\bar \z)(i + \z)} \right]^{-1/4}
   \right)
\nonumber
\eea      
\blank{
 z = x + I y;
 zb = x - I y;
 f = I  2^(-3/2) *  (2 y)^(-1/4) *  
  ( 
   (  (z-I)^(1/4)*(z+I)^(-1/4)*(zb+I)^(1/4)*(zb-I)^(-1/4)*
           ( 1/(z-1) - 1/(zb-1) + 1/(I-1) - 1/(-I-1)) )
   - 
   (  (z-I)^(-1/4)*(z+I)^(1/4)*(zb+I)^(-1/4)*(zb-I)^(1/4)*
           ( 1/(z-1) - 1/(zb-1) - 1/(I-1) + 1/(-I-1))  )
  );
 f = I  2^(-3/2) *  (2 y)^(-1/4) *  
  ( 
   (  (z-I)^(1/4)*(z+I)^(-1/4)*(zb+I)^(1/4)*(zb-I)^(-1/4)*
           ( 1/(z) - 1/(zb) + 1/(I) - 1/(-I)) )
   - 
   (  (z-I)^(-1/4)*(z+I)^(1/4)*(zb+I)^(-1/4)*(zb-I)^(1/4)*
           ( 1/(z) - 1/(zb) - 1/(I) + 1/(-I))  )
  );
fp3 = Plot3D[ f ,{x,-2.001,2},{y,0.01,2}, 
             PlotPoints -> 41 , ViewPoint -> {2,-1,1}
        ];
Display[ "fp3.psm" , fp3];
!psfix -epsf fp3.psm >! fp3.eps 
fp4 = Plot3D[ Sqrt[f] ,{x,-2.001,2},{y,0.01,2}, 
             PlotPoints -> 40 , ViewPoint -> {2,-1,1}, Mesh -> False
        ];
fp4 = Plot3D[ Abs[Sqrt[f]] ,{x,-2.,2.},{y,0.002,2}, 
             PlotPoints -> {61,40} , ViewPoint -> {2,-1,1}
        ];
graph4 = Show[ fp4, ViewPoint -> {1,-1,1} ];
fp4a = Plot3D[ Abs[Sqrt[f]] ,{x,-2.,-0.00001},{y,0.002,2}, 
             PlotPoints -> {30,40} , ViewPoint -> {2,-1,1} 
        ];
fp4b = Plot3D[ Abs[Sqrt[f]] ,{x,.000001,2},{y,0.002,2}, 
             PlotPoints -> {30,40} , ViewPoint -> {2,-1,1} 
        ];
graph4 = Show[ fp4a, fp4b , ViewPoint -> {1,-1,1} , PlotRange -> {-0.1,2}
        ];
Display[ "fp4.psm" , graph4 ];
!psfix -epsf fp4.psm >! fp4.eps 
fp5 = Plot3D[ - Sign[x] * Sqrt[f] ,{x,-2.001,2},{y,0.01,2}, 
             PlotPoints -> 40 , ViewPoint -> {2,-1,1}, Mesh -> False
        ];
fp5a = Plot3D[ - Sign[x] * Abs[Sqrt[f]] ,{x,-2.,-0.00001},{y,0.002,2}, 
             PlotPoints -> {30,40} , ViewPoint -> {2,-1,1} 
        ];
fp5b = Plot3D[ - Sign[x] * Abs[Sqrt[f]] ,{x,.00001,2},{y,0.002,2}, 
             PlotPoints -> {30,40} , ViewPoint -> {2,-1,1} 
        ];
graph5 = Show[ fp5a, fp5b , ViewPoint -> {1,-1,1} , PlotRange -> {-2,2}
            ];
Display[ "fp5.psm" , graph5];
!psfix -epsf fp5.psm >! fp5.eps 
}
This expression is single valued in $\z$, but when we take the
square root to find the result for a single Ising model there are
square-root branch points at $\z=0$ and at $\z=i$.
Where we put this branch cut 
has implications for the identification of the operators in the
correlation function, as shown in figure 3 where 
the functions plotted are identical up to a choice of the
position of the branch cut.

{
%\begin{figure}[ht]
\[
\ba{ccc}
  |f(\z)|^{1/2} 
= \vev{ \sigma(\z) \; \mu^B(0)\; \mu(i) }_{+}
&
& \!\!\!\!\!\!
  -{\rm sign}(\x)\, |f(\z)|^{1/2} 
= \vev{ \sigma(\z) \; \psi^{(+|-)}(0)\; \mu(i) }_{+|-}
\\
\epsfxsize=.45\linewidth
\epsfbox[0 40 288 268]{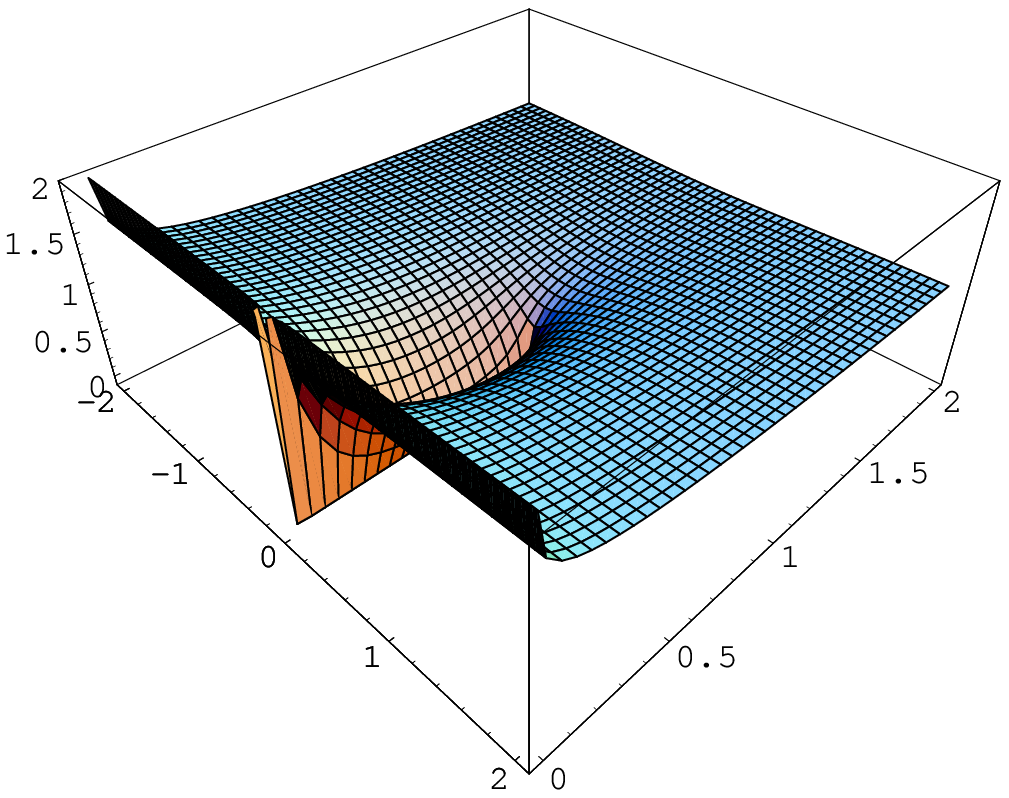}
%\epsfbox{++2.ps}
&
{\equiv}
&
\epsfxsize=.45\linewidth
\epsfbox[0 40 288 268]{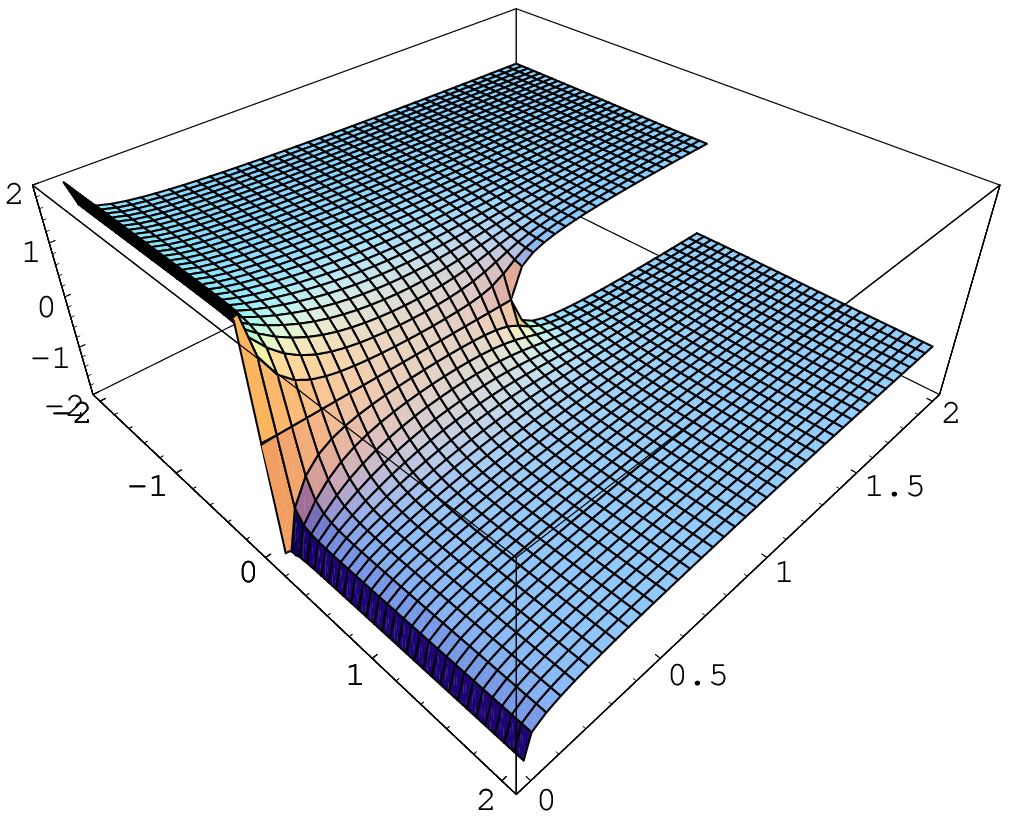}
%\epsfbox{+-2.ps}
\\[4mm]
\epsfxsize=.3\linewidth
\epsfbox{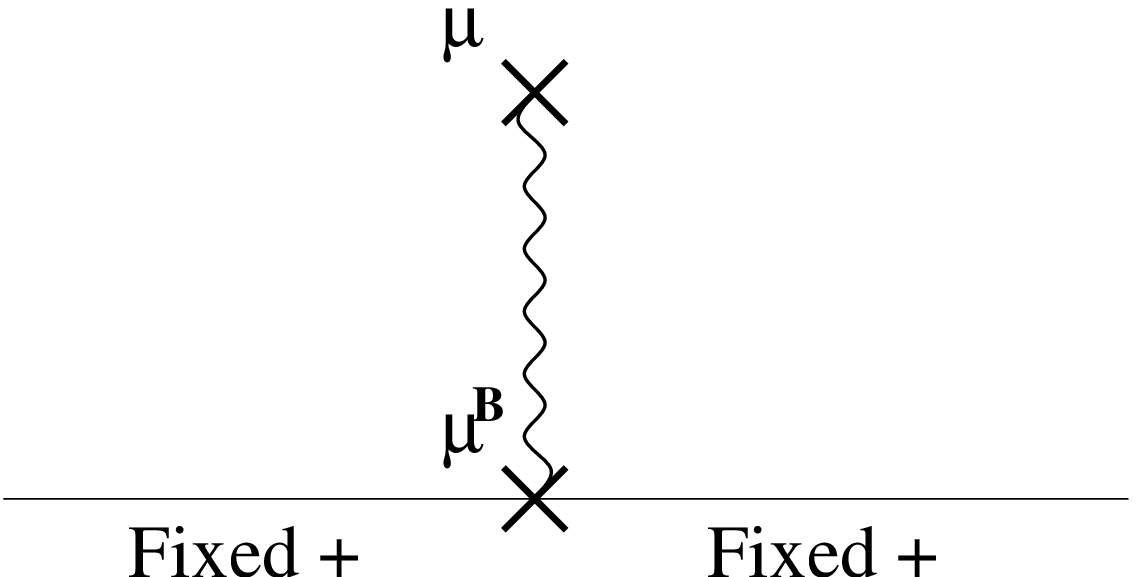}
&
&
\epsfxsize=.3\linewidth
\epsfbox{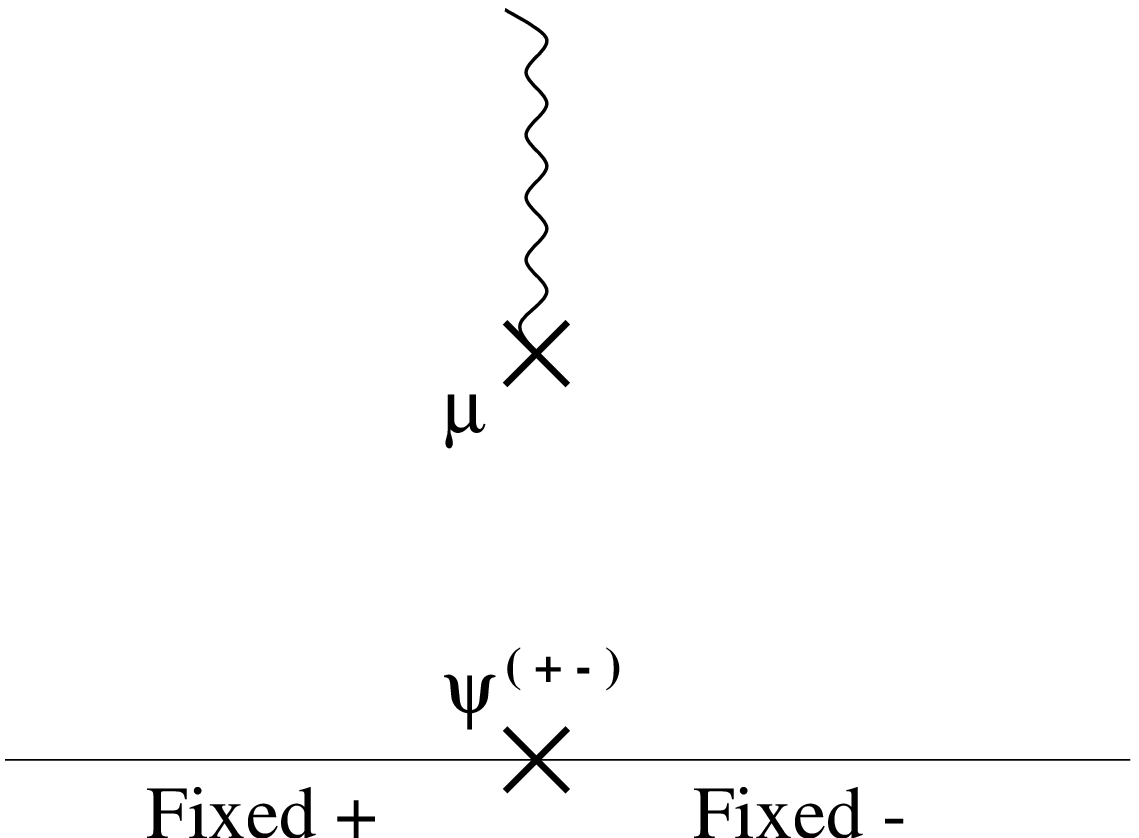}
\\[4mm]
\multicolumn{3}{c}{
\parbox[c]{.85\linewidth}
{\raggedright
Figure 3:
two choices of the position of the disorder line showing the
equivalent interpretation of the boundary disorder field $\mu^B$ as
the boundary-condition changing operator $\psi^{(\pm\mp)}$.
}
}
\ea
\]
%\caption{\small
%Figure \ref{fig:dps2}
%}
%\label{fig:fps2}
%\end{figure}
}

In the first case we plot $|f|^{1/2}$, and with this choice the
disorder line joins the bulk and boundary disorder fields, and the 
boundary condition appears to be uniform, fixed `up'.

In the second case we plot $-{\rm sign}(\x)|f|^{1/2}$, and the disorder
line from the bulk spin field extends to infinity in the bulk, and the
operator inserted at the origin appears to be a boundary
changing operator interpolating fixed 
`up' and fixed `down' b.c.'s.

% \newpage

\section{Conclusions}
\label{sec:conc}

We have shown how the bulk bosonisation formulae of Di Francesco et
al.\ \cite{FSZu1} can be extended to the Ising model on the upper half
plane when the bosonic field satisfies Dirichlet boundary conditions.
This has helped understand the nature of boundary disorder 
fields, and has also helped clarify the results of \cite{Ruelle} 
in which boundary states describing frustrated systems were found to
be rotationally invariant -- the result being that the disorder line
ends on an operator of weight zero and hence is invariant under rotations.

The boundary conformal field theory we have been looking at is unusual
on two counts. Firstly the bulk theory is non-local, and secondly it
seems we are forced to consider boundary conditions with apparently
two vacua; for the free case, these are vacua with even and odd
numbers of disorder lines, and in the fixed these are the fixed up and
fixed down vacua.  Such boundary conditions fall outside Cardy's
classification on both counts and it would be interesting to
understand more generally the relations between the non-locality of
the bulk theory and these peculiarities of the boundary theory. 

The perturbation by the boundary disorder operator also provides a
very simple example of a unitary perturbation by a boundary-condition
changing operator. Given the renewed interest in such perturbations
\cite{FGRS}, it will be of interest to generalise this to other
models. 

%\newpage

{\bf Acknowledgments}

I would like to thank K.~Graham, M.E.~Ortiz, A.~Recknagel, P.~Ruelle
and I.~Runkel for many helpful discussions, 
and the organisers of the 1999 Oberwolfach meeting on 
``Mathematical Aspects of String Theory''
for a very enjoyable meeting where this work was started.
The work was supported in part by a TMR grant of the European
Commission, contract reference ERBFMRXCT960012, 
and by an EPSRC advanced fellowship.

%%%%%%%%%%%%%%%%%%% BIBLIOGRAPHY %%%%%%%%%%%%%%%%%%%%%%

  \end{document}